\documentclass[]{interact}

\usepackage{epstopdf}% To incorporate .eps illustrations using PDFLaTeX, etc.
\usepackage{subfigure}% Support for small, `sub' figures and tables

\usepackage{natbib}% Citation support using natbib.sty
\bibpunct[, ]{(}{)}{;}{a}{}{,}% Citation support using natbib.sty
\usepackage{filecontents,pdfpages}

\theoremstyle{plain}% Theorem-like structures provided by amsthm.sty

\theoremstyle{definition}

\theoremstyle{remark}

\usepackage{hyperref}
\usepackage[utf8]{inputenc}
\usepackage{xcolor}
\def\tightlist{}
\usepackage{setspace}

 {% don't indent paragraphs
  \setlength{\parindent}{0pt}
  \ifodd #1 \everypar{\setlength{\hangindent}{\cslhangindent}}\ignorespaces\fi
  \ifnum #2 > 0
  \setlength{\parskip}{#2\baselineskip}
  \fi
 }%
 {}

\begin{document}

\articletype{}

\title{Hole or Grain? A Section Pursuit Index for Finding Hidden
Structure in Multiple Dimensions}

\author{\name{Ursula Laa$^{a, b, c}$, Dianne Cook$^{a}$, Andreas
Buja$^{d}$, German Valencia$^{b}$}
\affil{$^{a}$Department of Econometrics and Business Statistics, Monash
University; $^{b}$School of Physics and Astronomy, Monash
University; $^{c}$University of Natural Resources and Life Sciences,
Vienna, Department of Landscape, Spatial and Infrastructure Sciences,
Institute of Statistics, Peter-Jordan-Straße 82/I, 1190 Vienna,
Austria; $^{d}$Flatiron Institute, Simons Foundation, 162 Fifth Avenue,
New York, NY 10010, USA}
}

\thanks{CONTACT Ursula
Laa. Email: \href{mailto:ursula.laa@boku.ac.at}{\nolinkurl{ursula.laa@boku.ac.at}}, Dianne
Cook. Email: \href{mailto:dicook@monash.edu}{\nolinkurl{dicook@monash.edu}}, Andreas
Buja. Email: \href{mailto:andreasbuja@gmail.com}{\nolinkurl{andreasbuja@gmail.com}}, German
Valencia. Email: \href{mailto:german.valencia@monash.edu}{\nolinkurl{german.valencia@monash.edu}}}

\maketitle

\begin{abstract}
Multivariate data is often visualized using linear projections, produced
by techniques such as principal component analysis, linear discriminant
analysis, and projection pursuit. A problem with projections is that
they obscure low and high density regions near the center of the
distribution. Sections, or slices, can help to reveal them. This paper
develops a section pursuit method, building on the extensive work in
projection pursuit, to search for interesting slices of the data. Linear
projections are used to define sections of the parameter space, and to
calculate interestingness by comparing the distribution of observations,
inside and outside a section. By optimizing this index, it is possible
to reveal features such as holes (low density) or grains (high density).
The optimization is incorporated into a guided tour so that the search
for structure can be dynamic. The approach can be useful for problems
when data distributions depart from uniform or normal, as in visually
exploring nonlinear manifolds, and functions in multivariate space. Two
applications of section pursuit are shown: exploring decision boundaries
from classification models, and exploring subspaces induced by complex
inequality conditions from a multiple parameter model. The new methods
are available in R, in the \texttt{tourr} package.
\end{abstract}


\hypersetup{linkcolor=black}

\hypertarget{introduction}{%
\section{Introduction}\label{introduction}}

The visualization of high-dimensional data often utilizes linear
projection. For example, in principal component analysis plotting the
principal components is a projection of the data. Projections form the
basis for a grand tour \citep{As85} of high-dimensional data. Projection
pursuit \citep[\citet{FT74}]{kr69} is used to find interesting
low-dimensional views of the data by optimizing an index function over
all possible projections. Projections can obscure non-uniform patterns
near the data center, or hollowness. These features may be visible in
non-linear mappings (such as multidimensional scaling \citep{mds}), but
these methods often lack interpretability. \citet{laa2019slice}
introduced a slice tour that shows sections through high-dimensions
instead of projections, and helps to reveal hidden structure. Showing
sections can also be useful in combination with conditional function or
model visualization \citep[\citet{sliceplorer}]{JSSv081i05}.

The space of sections of high-dimensional space is larger than the space
of projections. The slice tour is based on projections, considering an
observation to be in the slice if it is within a fixed distance of the
projection plane, thus simplifying the space to be explored. This
definition of a slice allows for the comparison of distributions of
points inside and outside the slice. This is an unguided process, and
providing a method to find interesting slices would be beneficial.

Here we propose ``section pursuit'', which searches the space of
projections for the most interesting slices of the data. (Technically,
``slice pursuit'' might better match the methods proposed in this paper,
but ``section pursuit'' is preferred because it is more general and
applicable to a broader class of problems.) The ``interestingness'' of
each slice is computed as a measure of dissimilarity between the
distribution of projected points inside and outside the slice. Finding
conditional features in large data sets will provide better
understanding of the data, and can help improve modelling.

This paper is organized as follows. The next section provides background
on tours, slicing and projection pursuit. Section \ref{sec:index}
describes an index designed to detect concavities (holes), and its
reverse to detect grains. Section \ref{sec:practical} discusses the
associated practical considerations and explores the behavior of the
index on simulated examples. The application of section pursuit to
exploring high-dimensional classification models and constrained
high-dimensional spaces is illustrated in Section
\ref{sec:applications}.

\hypertarget{background}{%
\section{\texorpdfstring{Background
\label{sec:background}}{Background }}\label{background}}

\hypertarget{grand-tour-and-slice-tour}{%
\subsection{Grand tour and slice tour}\label{grand-tour-and-slice-tour}}

The grand tour shows a geodesically interpolated sequence of randomly
selected projections in an animation. The interpolation allows the
viewer to interpret each view in the context of the previously seen
projection, thus providing additional insights compared to static
projections. By observing how the data distribution changes under a
rotation of the low-dimensional projection, the viewer extrapolates from
the low-dimensional shapes to the distribution in high dimensions. The
underlying computational methods were described in \citet{BCAH05}, and
an implementation in R \citep{rlang} is available in the \texttt{tourr}
package \citep{tourr}.

\begin{figure*}[ht]
%\centerline{\includegraphics[width=0.4\textwidth]{diagrams/centered-slice.pdf}
%\hspace{5mm}
%\includegraphics[width=0.4\textwidth]{diagrams/orthogonal-distance.pdf}}
\centerline{\includegraphics[width=0.8\textwidth]{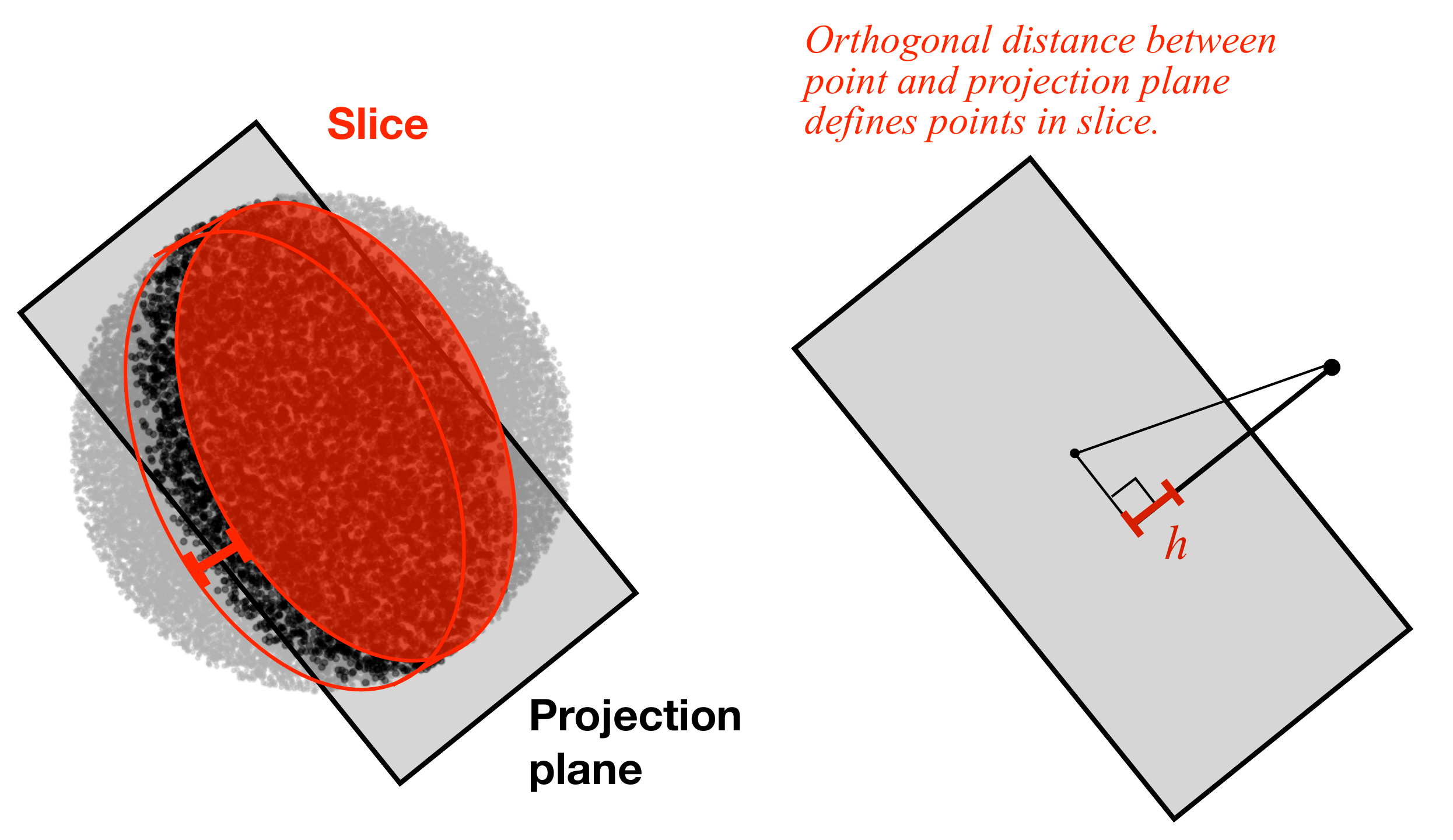}}
\caption{Illustrations of slicing, through a 3D sphere (left), and demonstrating the calculation of the orthogonal distance (right). A slice can be done centered at the origin or off-center. Orthogonal distance between point and projection plane is used for computing the slice.}
\label{fig:diagrams}
\end{figure*}

The information available in projections can be complemented using
\textit{sectioning} \citep{prosection}. Sections highlight (or ignore)
subsets of the data based on inequality conditions defined for the
variables. We use the term \textit{slicing} for a special sectioning
condition, that uses the orthogonal distance of data points from a
projection plane. Points are considered to fall into the slice if this
distance is below some cutoff value \(h\). Figure \ref{fig:diagrams}
shows diagrams illustrating slicing through a 3D sphere (left) and the
general concept of the orthogonal distance (right) that is used to
define slices in more than three dimensions. Recent work in
\citet{laa2019slice} implemented a display of interpolated slices, which
has been added to the \texttt{tourr} package and can be used to show
slices for projection planes obtained when running a grand tour.

A projection plane does not generally need to have a location associated
with it, but it is necessary in order to compute the distance between a
point and the plane. By default, the projection plane is considered to
pass through the center of the data, or \(\boldmath{0}\) if the data has
been centered. Using Euclidean distance in the orthogonal space is done
to ensure rotation invariance. This raises an interesting geometric
result, that the slices are spherical, one might even call them shells.
A \textit{flat slice}, matching that shown in Figure
\ref{fig:diagrams}), is only obtained in the case of a single orthogonal
direction on the projection plane (i.e.~3D). However, the appearance to
the viewer watching the slice tour is that of a slice rather than a
shell. Hence, the use of the the term slice.

\begin{figure}

{\centering \includegraphics[width=0.8\linewidth]{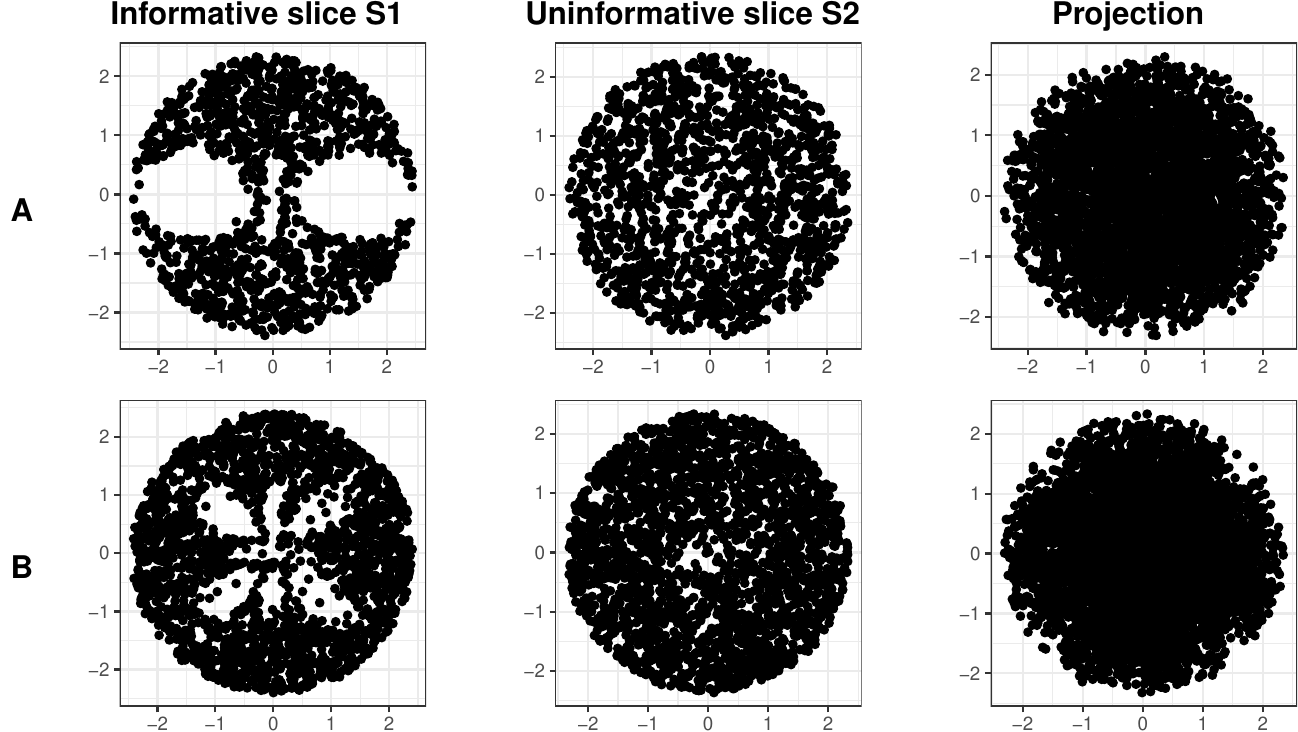} 

}

\caption{Slices of two 4D data sets (A, B) shown using scatterplots of points inside the slice. The first slice (S1) is informative and reveals the hollow regions in the data, while the second slice (S2) is un-informative. The last column shows projections of the data onto a 2D plane, hiding the hollowness.}\label{fig:allSlices}
\end{figure}

The first two columns of Figure \ref{fig:allSlices} show scatterplots of
the points in a 4D data set inside an informative (S1) and an
uninformative (S2) slice, defined by two different projection planes
centered at the origin, for two simulated data sets (A, B). Note that
the data is defined in a 4D sphere, which generates the circular shape.
The sampling results in hollow regions inside the sphere by rejecting
points within shapes defined by selected hyperspherical harmonics
\citep{doi:10.1063/1.3054274}, as described in the Appendix. The
hollowness is hidden in projections of the data, as shown in the final
column of Figure \ref{fig:allSlices}. It is much simpler to define
slices on high-dimensional spheres than cubes, and this is the approach
used for all the data in this paper. Data defined in a cube is trimmed
or supplemented to be contained in a sphere, which we would not expect
to lose important information because the focus is on detecting hidden
features in the center of the distribution.

\hypertarget{projection-pursuit-and-guided-tour}{%
\subsection{Projection pursuit and guided
tour}\label{projection-pursuit-and-guided-tour}}

Projection pursuit is the procedure of selecting interesting
low-dimensional projections by optimizing a criterion function over the
space of all possible projections. The criterion (or index) function is
typically designed to take larger values for more ``interesting'' views
of the data, and maximizing the index can reveal structure in the
distribution.

The concepts of projection pursuit and the grand tour can be combined
into the guided tour \citep{CBCH94}, which interpolates between
projection planes selected through the optimization of a projection
pursuit index. A guided tour presents the viewer with interesting views
of the dataset in the context of the full distribution, moving from less
to more informative projections.

Tour methods can also be used to understand the behavior of projection
pursuit index functions \citep{laa2020}, for example by examining how
the index value changes along an interpolated sequence of projections.

\hypertarget{a-new-index-for-finding-interesting-sections}{%
\section{\texorpdfstring{A new index for finding interesting sections
\label{sec:index}}{A new index for finding interesting sections }}\label{a-new-index-for-finding-interesting-sections}}

We define a new index function based on comparing the distribution of
points inside a slice to the distribution outside the slice. This is
achieved by comparing normalized counts in bins across the projection
plane to find differences between the distributions. We first introduce
the notation for slicing and binning (3.1), followed by the definition
of the index function itself (3.2), and assumptions to be made are
discussed in the next section.

\hypertarget{taking-a-slice-and-binning}{%
\subsection{Taking a slice and
binning}\label{taking-a-slice-and-binning}}

Let \(Y=X\cdot A\), where \(X\) is an \(n\times p\) data matrix, \(A\)
is a \(p\times d\) (orthonormal) basis for the \(d\)-dimensional space
onto which the data is being projected. To generate a 2-dimensional
slice, following \citet{laa2019slice}, compute the orthogonal distance
between every point and the plane (defined by
\(A = (\mathbf{a}_1, \mathbf{a}_2)\)) as the Euclidean norm
\begin{equation}
h_i = ||\mathbf{x}_i - (\mathbf{x}_i\cdot \mathbf{a}_1) \mathbf{a}_1 - (\mathbf{x}_i\cdot \mathbf{a}_2) \mathbf{a}_2||.
\end{equation} Observations are considered inside the slice if
\(h_i < h\).

We denote the set of points inside the slice \(S\), and the set of
points outside the slice \(C\). (You can think about \(C\) being
short-hand for \(S^C\), the complement of \(S\).) For the definition of
the index we will further bin the projected data \(Y\) into \(K\) bins,
\(b_k, k=1, ..., K\). The counts of observations for each bin are
computed separately for \(S\) and \(C\). We can thus write the counts in
bin \(k\) via two indicator functions,
\(S_{k}=\sum_{i} I(Y_i \in b_{k})I(h_i < h)\) inside the slice and
\(C_{k}=\sum_{i} I(Y_i \in b_{k})I(h_i \geq h)\) outside the slice. The
relative counts are thus \(s_k = S_k / \sum_i S_i\) and
\(c_k = C_k / \sum_i C_i\).

Data points are therefore classified in two ways: information about the
projected position onto the viewing plane is captured by the binning,
while information about the position in the orthogonal space is captured
by slicing. Thus the first step is projecting and slicing, followed by a
separate binning of the projected points \(Y\) for the subsets \(S\) and
\(C\).

\hypertarget{index-definition}{%
\subsection{Index definition}\label{index-definition}}

Indexes for comparing the distribution of observations in the slice
versus outside can be defined (building on
\citet{doi:10.1198/1061860043119}) as:

\begin{equation}
I_A^{low} = \sum_{k}\left[\left(c_{k}-s_{k}\right)\right]_{>\varepsilon},
\label{eq:index}
\end{equation}

\begin{equation}
I_A^{up} = \sum_{k}\left[\left(s_{k}-c_{k}\right)\right]_{>\varepsilon}
\label{eq:indexup}
\end{equation}

\noindent where \(s_{k}\) and \(c_{k}\) are the relative counts of
observations in bin \(b_{k}\) inside and outside the slice. The first
definition \(I_A^{low}\) takes large values if there is a hollow region
of low density inside the slice (hole), while the second definition
\(I_A^{up}\) can be useful when looking for regions of higher density
(grain or possibly considered to be ``needles in a haystack''). The
\([.]_{>\varepsilon}\) notation indicates that we drop all bins where
the difference in counts is below some threshold \(\varepsilon\),
\begin{equation}
[a - b]_{>\varepsilon} = \begin{cases}
    a - b, & \text{if $a - b > \varepsilon$}\\
    0, & \text{if $a-b \leq \varepsilon$.}
  \end{cases}
\end{equation} The \(\varepsilon\) avoids summing noise and suppresses
an artificial dependence on the number of bins. The value of
\(\varepsilon\) should be estimated based on the expected size of
sampling fluctuation, which may depend on assuming a distribution, the
number of samples \(N\), the number of bins \(K\), the slice radius
\(h\) and the dimension \(p\). An estimate for \(\varepsilon\) assuming
a uniform distribution inside a hypersphere is given in Section
\ref{sec:epsilon}. If \(\varepsilon=0\) we get a symmetry in the
definitions, \(I_A^{low}=I_A^{up}\). In general, the choice of index
should match the intention, searching for regions of high (grain) or low
(hole) density in the slice.

\hypertarget{sec:generalise}{%
\subsubsection*{Generalised index}\label{sec:generalise}}
\addcontentsline{toc}{subsubsection}{Generalised index}

The definitions in Eq.(\ref{eq:index},\ref{eq:indexup}) can be
generalized for example to emphasize selected bins or control the
sensitivity. \begin{equation}
I_A^{low} = \sum_{k}w_{k}\left(\left[c_{k}^{1/q}-s_{k}^{1/q}\right]_{>\varepsilon}\right)^{q}
\label{eq:index2}
\end{equation} \begin{equation}
I_A^{up} = \sum_{k}w_{k}\left(\left[s_{k}^{1/q}-c_{k}^{1/q}\right]_{>\varepsilon}\right)^{q}.
\label{eq:index2up}
\end{equation}

Here \(w_k\) can be used to (de)emphasize certain bins, e.g.~to
up-weight information in the center of the distribution. The exponent
\(q\) can be used to tune the sensitivity, a small \(q\) will enhance
sensitivity to small differences (and thus might over-emphasize
fluctuations), a large \(q\) will suppress fluctuations and is mainly
sensitive to large differences (and might thus miss features that are
too similar to the background distribution). Selecting \(q=1\)
corresponds to an \(L_1\) type norm, and \(q=1/2\) to an \(L_2\) type
norm. In addition we will also consider \(q=2\).

The overall range of plausible index values depends on \(q\) and in
practice we use an estimate of the range to rescale the index value to
fall in \([0,1]\). Notice that, in this definition the index is no
longer symmetric under exchanging the distributions inside and outside
the section, even when \(\varepsilon=0\).

\hypertarget{practical-issues}{%
\section{\texorpdfstring{Practical issues
\label{sec:practical}}{Practical issues }}\label{practical-issues}}

The application of the new index function should be informed by an
expected underlying distribution. As explained in \ref{sec:rotinv} these
should be rotation invariant, and the concrete assumption can guide
practical choices that need to be made: reweighting bin counts to
account for expected differences between the projected distribution and
the distribution in a slice (\ref{sec:binning}), estimates of sufficient
sample size (\ref{sec:size}), expected noise and corresponding threshold
\(\epsilon\) (\ref{sec:epsilon}). We also include visualizations to
better understand the index behavior and how it depends on its
parameters (\ref{sec:behaviour}) and guidance on using the index in
practice (\ref{sec:viz}).

\hypertarget{rotation-invariance}{%
\subsection{\texorpdfstring{Rotation invariance
\label{sec:rotinv}}{Rotation invariance }}\label{rotation-invariance}}

A desirable property for projection pursuit, into \(d\)-dimensions, and
thus also desirable for section pursuit, is that the index be rotational
invariant. That is, regardless of the basis in the \(d\)-D plane
defining the projection, the index value should be identical. With
slices, this is more complicated, because interior and exterior
distributions need to be comparable in the absence of structure. To
account for expected differences we need to choose an assumed
(spherically symmetric) distribution of the data.

Thus a restriction is imposed on the observed data: that it falls within
a \(p\)-dimensional hypersphere. Data is typically observed in a
hypercube, so this prescription requires a departure from the
convention. However, it is still practical, and reasonable. For example,
when using simulated data to examine multivariate models, like
classification boundaries, the sampling scheme can produce points in a
hypersphere upon which to make model predictions. On the other hand,
observed data may require shaving off the corners. Because we are
interested in interior structures near the center of the distribution
the corners are less important, and we do not expect to lose relevant
information. Moreover, projections of several variables have a tendency
to produce elliptical or spherical shapes (\citet{diaconis1984} pointed
out that most projections are approximately Gaussian, and
\citet{burningsage} showed that the effect of piling near the center is
sizable already with moderate number of dimensions, \(p\approx10\)).
This can also be assisted by sphering the data during pre-processing. In
addition, this approach is designed for reasonably small \(p\) so that
the vast gap between spheres and cubes in high dimensions is not a
concern. For very high-dimensional data, some dimension reduction is a
necessary part of pre-processing.

Here we will work with the assumption of a uniform distribution inside a
hypersphere because it naturally captures how the relative slice volume
changes with dimensionality, that is, it reflects that we are operating
in Euclidean space. Other rotation invariant data distributions, for
example, multivariate normal, could also be of interest as a reference.

\hypertarget{polar-binning-and-reweighting}{%
\subsection{\texorpdfstring{Polar binning and reweighting
\label{sec:binning}}{Polar binning and reweighting }}\label{polar-binning-and-reweighting}}

Rotation invariance is a desirable feature for a section (or projection)
pursuit index and this suggests a preference for binning in polar
coordinates. We can decompose an expected underlying distribution into a
radial and a directional component, the latter being parametrized by
\(d-1\) angles. The expected reference distribution can then guide the
good choices about binning on these parameters.

\begin{figure}

{\centering \includegraphics[width=0.3\linewidth]{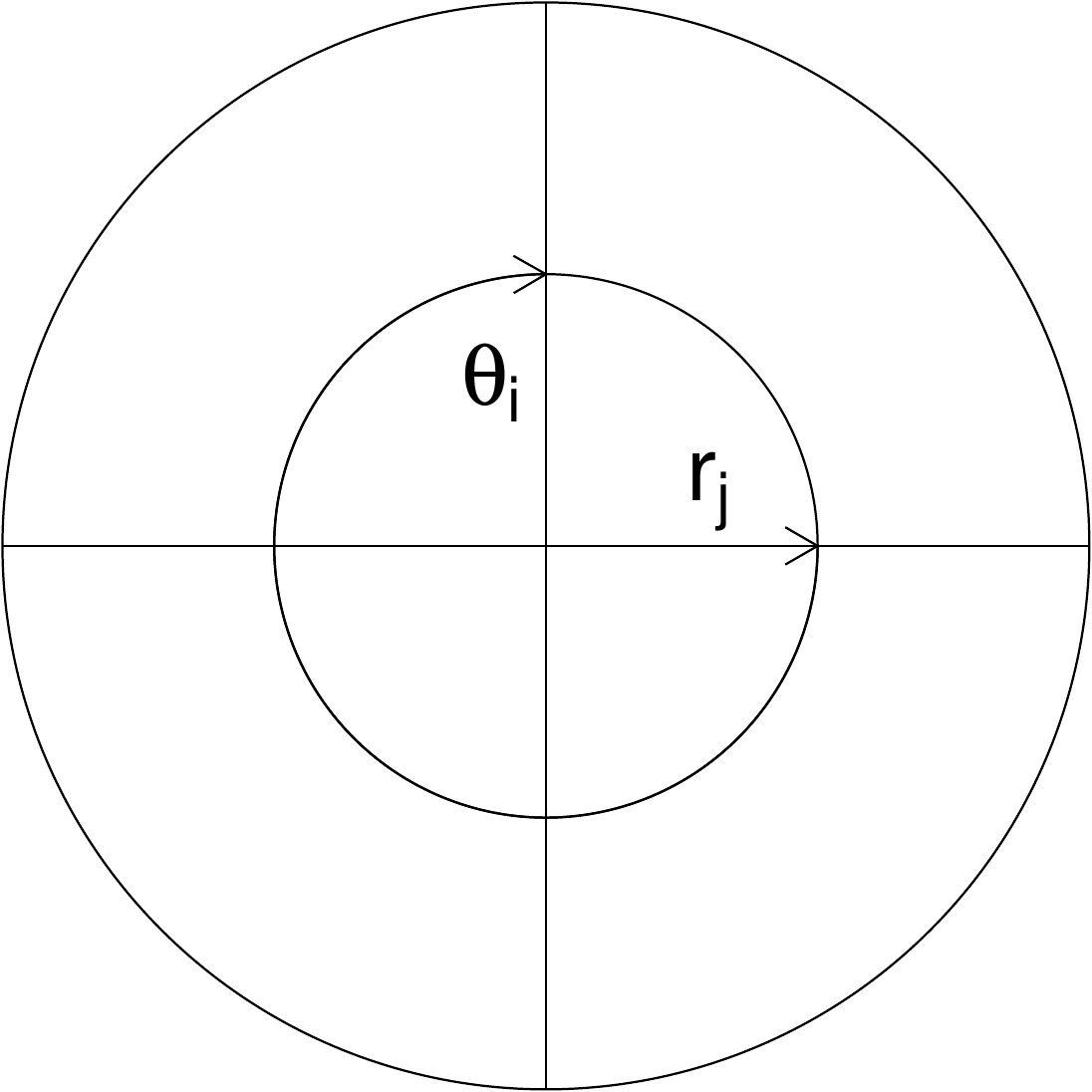} 

}

\caption{Diagram illustrating the notation for polar binning in angle $\theta$ and radius $r$. Angular bins $\theta_i, i=1, 2, 3, 4$, split the angular component into four bins, $r_j, j=1, 2$ splits the radial component into two bins, resulting in a total of 8 bins.}\label{fig:sketch}
\end{figure}

In our case we are interested in binning the projected data points in
\(d=2\) dimensions and will use the radius \(r\) and a single angle
\(\theta\) in the projection plane to parametrize the binning, as
illustrated in the diagram in Figure \ref{fig:sketch}. As a reference
distribution we consider points that are uniformly distributed in a
hypersphere, thus the angular distribution will be uniform across all
values of \(\theta\), suggesting the use of \(K_{\theta}\) equidistant
angular bins. This would also hold if the reference distribution was
assumed to be a spherical distribution (e.g.~the product of i.i.d.
univariate normals).

The radial binning is more complicated. This is because the marginal
radial distribution of the projected data depends on the assumed
reference distribution as well as the data dimension \(p\). As \(p\)
increases, the projected data piles more in the center, for an assumed
spherical uniform or normal reference distribution \citep{burningsage}.
A varying radial bin size can be used to offset this effect, where the
bounds of the \(K_r\) radial bins take into account the reference
distribution of points given \(p\). However, the expected radial
distribution will also differ between the projected data in a slice and
a projection of the full data, so the effect is accounted for instead by
reweighting bin counts. The aim is to adjust for the expected difference
in distribution: after the reweighting, the expected counts in all
radial bins is the same, and this needs to be adjusted separately for
the data in the slice and the projected data. The calculations are as
follows.

\begin{figure}

{\centering \includegraphics[width=0.5\linewidth]{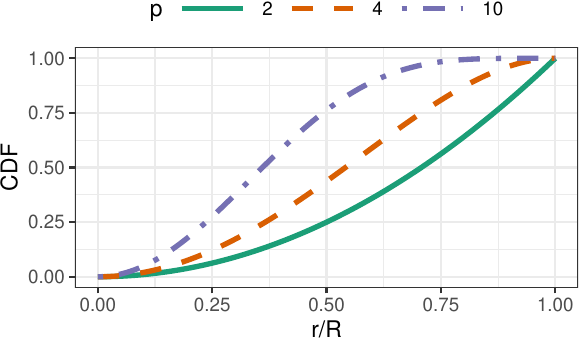} 

}

\caption{CDF of the 2D projected radial distribution of a p-dimensional hypersphere, for selected values of p. As p increases, a larger fraction of the points will be found at small values of the radius.}\label{fig:cdf}
\end{figure}

Consider the expected distribution to be a uniform distribution within a
hypersphere in \(p\) dimensions projected onto a 2D plane. The
cumulative distribution function (CDF) for the radial distribution in
the projection, derived in the Appendix, is given by:

\begin{equation}
F(r;p,R) = 1-\left(1-\left(\frac{r}{R}\right)^2\right)^{p/2},
\label{eq:cdf}
\end{equation} and depends on the hypersphere radius \(R\) and on \(p\).
For illustration we show the CDF dependence on \(r/R\) (where
\(0<r\leq R\)) for different values of \(p\) in Figure \ref{fig:cdf}.
For large values of \(p\) the majority of points is found within a small
relative radius \(r/R\), for example for \(p=10\) we see that 75\% of
points are within \(r/R<1/2\).

While the radial distribution of the points projected from a full \(p\)
dimensional hypersphere follows the CDF in Eq.(\ref{eq:cdf}), the
distribution in the slice can be made to be approximately uniform in the
disk for some \(h \ll R\). Within the slice, the adjustment only
accounts for relative areas in radial bins for a 2D uniform
distribution. All weights can be calculated from the CDF.

The fraction of points in the radial bin \(r_i\) with lower boundary
\(r_i^l\) and upper boundary \(r_i^u\) is \begin{equation}
F_{i}(p, R) = F(r_i^u; p, R) - F(r_i^l; p, R).
\label{eq:frac}
\end{equation} Consider \(K_r\) radial bins, with total bin count
\(N_i\) in bin \(i\), \(i = 1,...,K_r\), and relative counts are
\(n_i = N_i / \sum_j N_j\). We define the bin-wise weights as
\begin{equation}
w_i(p, R, K_r) = \frac{1}{K_r F_i(p,R)},
\end{equation} and the reweighted bin count as \begin{equation}
s_i(p, R, K_r) = n_i w_i(p, R, K_r)
\label{eq:reweight}
\end{equation} corresponding to the weighted relative number of points
in each bin. After reweighting, the relative expected count in each
radial bin is thus \(1/K_r\) for a uniform hypersphere. We calculate the
outside weights as \(w_i(p, R, K_r)\) and the inside weights as
\(w_i(2, R, K_r)\). Recall that we are assuming that the slice thickness
can be neglected, \(h\ll R\).

\begin{figure}

{\centering \includegraphics[width=0.9\linewidth]{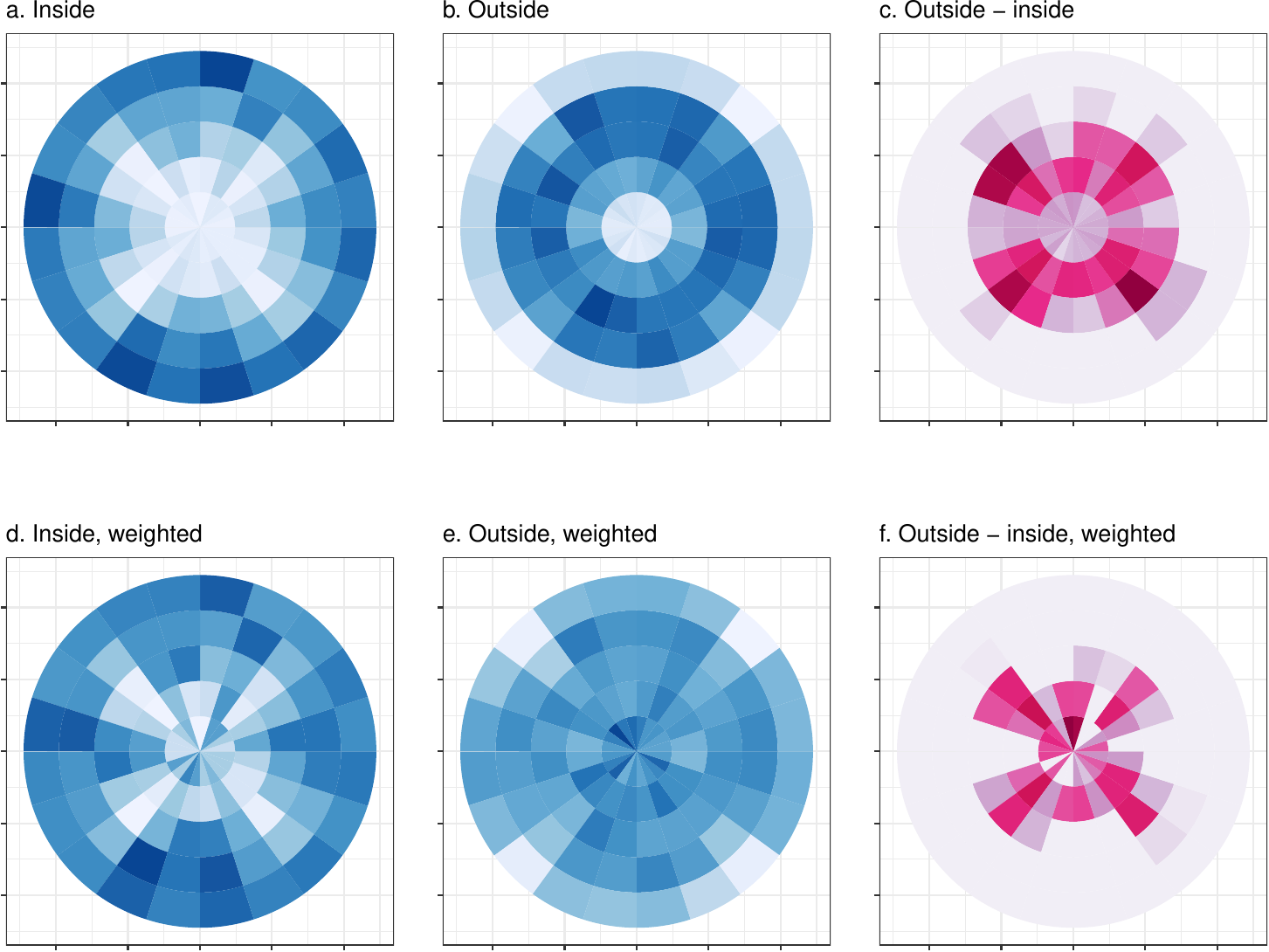} 

}

\caption{Illustration of the polar binning, effect of re-weighting bin counts in the index function computation, on sample data set B and for the informative slice S1. Rows show raw and weighted bin counts, respectively. Columns show the distributions of inside, outside and the difference. The effect of the re-weighting focuses the attention more effectively on the cavity structure.}\label{fig:data_and_densities}
\end{figure}

Figure \ref{fig:data_and_densities} illustrates the effect of the
adjustment using polar histograms for the simulated example data B, with
and without re-weighting. The rows show raw and weighted bin counts,
respectively, and the columns contain plots of the distributions of
points inside, outside the slice and the difference between the two. The
re-weighting has the desired effect of focusing the attention more
effectively on the central structure.

The index needs to be optimized over the set of all possible projections
to find interesting concavities. This requires the binned data to be
comparable, thus \emph{the bins are fixed for all planes}. For each
slice we first center the projected data before binning it.

In the simulated examples in this paper the radius of the hypersphere,
\(R\), is known, and is used to define the radial boundary. In practice,
the radius of a random projection can be used to estimate \(R\). When
observed points for any new projection, or slice, fall outside of the
sphere, they might be grouped into the outer bin.

Note that, the fixed angular binning can marginally affect rotational
invariance of the index. To mitigate this, combining index values from
several small rotations within a single angular bin window might be done
at each step.

\hypertarget{sufficient-sample-size}{%
\subsection{\texorpdfstring{Sufficient sample size
\label{sec:size}}{Sufficient sample size }}\label{sufficient-sample-size}}

As dimensionality increases the number of sample points required to
resolve features in a thin slice of the data increases exponentially.
Care must be taken that the sample is large enough. We estimate the
number of required sample points as a function of the chosen parameters,
starting again from the hyperspherical distribution in \(p\) dimensions.
Following \citet{laa2019slice}, given a sample of \(N\) points in \(p\)
dimensions, distributed uniformly in a hypersphere of radius \(R\), the
number of points inside a slice through the center of the distribution
is

\begin{equation}
N_S = \frac{N}{2} \left(\frac{h}{R}\right)^{p-2} \left(p - (p-2)\left(\frac{h}{R}\right)^{2}\right).
\label{eq:count}
\end{equation}

\noindent The calculation is based on the relative volume of the slice,
derived in the Appendix. We denote \(x=h/R\) the resolution. It
determines the minimum relative size of features that can be seen in a
slice. When adding noise dimensions, in order to keep the resolution
\(x\) fixed, the sample size needs to increase approximately as
\(x^{\Delta p}\), where \(\Delta p\) is the increase in number of
dimensions.

As an explicit example consider the expected fraction of points
\(N_S/N\) inside a slice of resolution \(x=\frac{h}{R}=0.1\). This
fraction is about \(0.15\) or \(15\%\) when \(p=3\), and quickly drops
to \(0.02\) when \(p=4\) and \(0.002\) when \(p=5\). Thus, even for
moderate dimensionality \(p\) the original sample size needs to be large
to have enough points to resolve features in a slice.

\hypertarget{estimating-the-magnitude-of-noise-varepsilon}{%
\subsection{\texorpdfstring{Estimating the magnitude of noise,
\(\varepsilon\)
\label{sec:epsilon}}{Estimating the magnitude of noise, \textbackslash varepsilon }}\label{estimating-the-magnitude-of-noise-varepsilon}}

We can estimate the expected sampling variability based on \(N\). The
dominant uncertainty will be on the bin count inside the slice which
typically will have much smaller statistics than the outside
distribution (notice that this may not be true for bins at large radius
which, depending on the bin size, can have a very low number of
observations). We estimate the number of points in a bin \(i\) inside
the slice as \begin{equation}
N_S^i = \frac{N_S}{K_{\theta}} \cdot F_i(2, R),
\end{equation} with \(N_S\) given by Eq.(\ref{eq:count}), \(K_{\theta}\)
the number of angular bins and \(F_i\) defined in Eq.(\ref{eq:frac}).
The relative Poisson error on this count is \begin{equation}
\delta_S^i = \frac{\sqrt{N_S^i}}{N_S^i} = 
\frac{R}{\sqrt{(r_i^u)^2 - (r_i^l)^2}} \sqrt{\frac{2 K_{\theta}}{N}} x^{(2-p)/2}
\frac{1}{\sqrt{\left(p - (p-2) x^{2}\right)}}, 
\end{equation} where \(r_i^l\) and \(r_i^u\) are the lower and upper
radial boundary of bin \(i\) and the expected error will be different
depending on the radial position of the bin. By its definition, the
expected reweighted count is \(1/K\) in all bins, where
\(K=K_r K_{\theta}\). We therefore expect sampling fluctuations of order
\begin{equation}
\delta^i = \delta_S^i / K. 
\label{eq:eps}
\end{equation}

\begin{figure}

{\centering \includegraphics[width=0.9\linewidth]{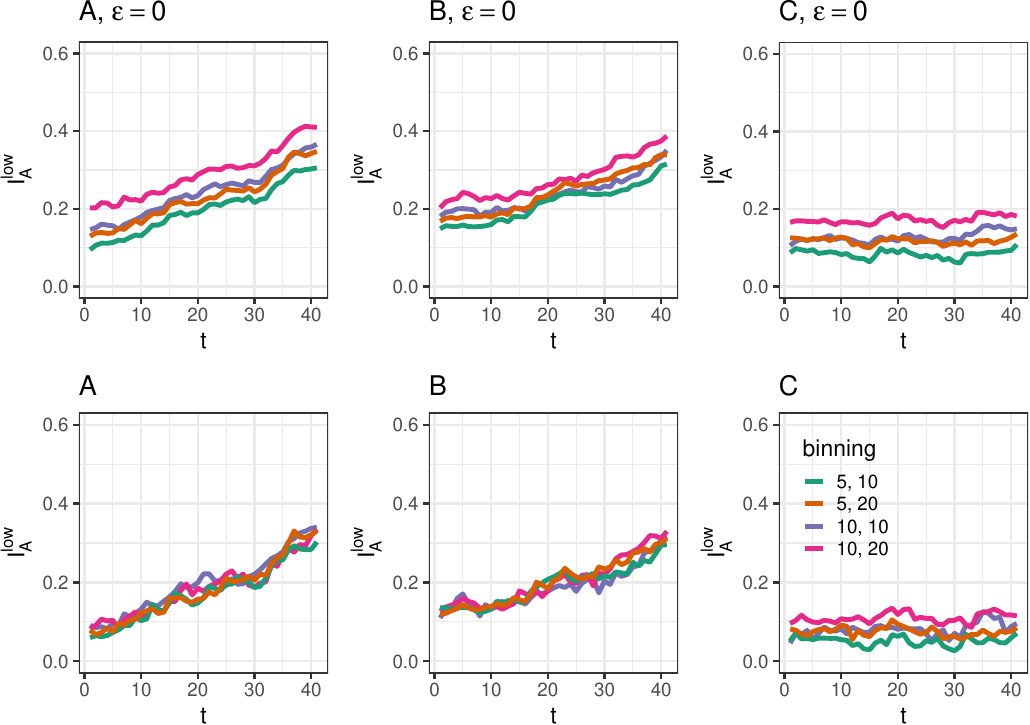} 

}

\caption{Examining the effect of the noise cutoff $\varepsilon^i$ on the index value for varying number of bins. Three simulated data sets are used:  A, B and C. The horizontal axis traces the index from an uninteresting to interesting sliced projection (defined by slice S2 and S1 respectively), for $\varepsilon^i=0, \delta^i$. Colour indicates number of bins. Ideally, the index value should be the same, regardless of choice of bins, which is achieved in practice by $\varepsilon^i=\delta^i$.}\label{fig:path_eps}
\end{figure}

To suppress index fluctuations to below one standard deviation, we set a
bin-wise \(\varepsilon^i = \delta^i\). Figure \ref{fig:path_eps}
compares the index behavior for two values of \(\varepsilon^i\), \(0\)
and \(\delta^i\). Three 4D data sets are used: examples A and B (shown
in Figure \ref{fig:allSlices}), and a reference set C consisting of
observations sampled uniformly within a \(p\)-dimensional sphere. Color
indicates the number of bins used to calculate the index. The difference
in index value when \(\varepsilon^i=0\) for all bins clearly indicates a
dependence on the number of bins, which is undesirable. Setting
\(\varepsilon^i = \delta^i\) mostly removes these differences.

\hypertarget{dependenceq}{%
\subsubsection*{\texorpdfstring{Dependence on
\(q\)}{Dependence on q}}\label{dependenceq}}
\addcontentsline{toc}{subsubsection}{Dependence on \(q\)}

\begin{figure}

{\centering \includegraphics[width=0.5\linewidth]{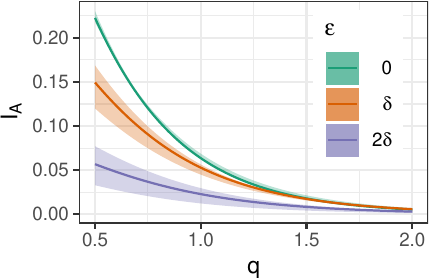} 

}

\caption{Examining the  effect of the power parameter $q$ on the variability of the  index value, when a noise cutoff ($\varepsilon$) is used. Rescaled values of $I_A$ shown as a function of $q$, for fixed values of $\varepsilon$, for $N=10000$ and $K=100$. Smaller values of  $q$ inflate the variability in the index, exaggerating  the noise.}\label{fig:noise}
\end{figure}

In the generalized definition of the index, the parameter \(q\) can be
used to tune the sensitivity to small differences. It will therefore
also need to be considered when choosing \(\varepsilon^i\). To
understand the effect we consider a simple one-dimensional setup:
uniformly draw \(N\) samples and bin them in \(K\) bins, such that the
expected bin count is \(N/K\) in all bins (or \(1/K\) after
normalization). Using the Poisson approximation, the standard deviation
for the normalized counts is \(\delta=1/\sqrt{N K}\). By drawing two
independent samples and calculating the index value 100 times we get an
estimate of the expected index value and its variance depending on
\(q\). Note that since we are assuming pure noise distributions for both
samples, \(I_A^{low}=I_A^{up}=I_A\). This is shown in Figure
\ref{fig:noise} for \(N=10000\) and \(K=100\). We see that for values of
\(q<1\) the noise becomes inflated, especially when we do not use an
appropriate \(\varepsilon\) cutoff. For the default choices of \(q=1\)
and \(\varepsilon=\delta\) we find that the expected index value for a
noise distribution is about \(0.05\). Note that in this simplified setup
there is no bin-dependence of \(\varepsilon\) and we have dropped the
index \(i\).

\hypertarget{index-behavior}{%
\subsection{\texorpdfstring{Index behavior
\label{sec:behaviour}}{Index behavior }}\label{index-behavior}}

To examine the behavior of the generalized index for choices of \(q\),
\(K\) and \(\varepsilon\), a re-parametrization is used. Consider an
outside distribution that is uniform across all \(K\) bins,
\begin{equation}
c_{k} = \frac{1}{K},
\end{equation} and a distribution inside the slice to be uniform across
\(K' < K\) bins, \begin{equation}
s_{k} = \begin{cases}
    \frac{1}{K'}, & \text{if $k \leq K'$}\\
    0, & \text{if $k > K'$}
  \end{cases}
\end{equation} that is there are \(K-K'\) empty bins inside the slice.
We can then parametrize the relation between \(K\) and \(K'\) as
\(K' = \gamma K\) with \(\gamma \in [0,1]\).

To focus on the dependence on \(q\), drop the bin-dependent weights
\(w_k\) from the description and set \(\varepsilon=0\). Using the
parametrization in terms of \(\gamma\), the index can then be written as

\begin{equation}
I_A^{up}(q, \gamma) = ([1-\gamma^{1/q}]_+)^q,
\label{eq:p1}
\end{equation}

\noindent while \begin{equation}
I_A^{low}(\gamma) = 1-\gamma.
\end{equation} (In this latter form the dependence on \(q\) drops out
because only bins where \(s_k=0\) are counted.)

This allows us to estimate the typical range depending on \(q\). The
range of \(I_A^{up}(q, \gamma)\) will be different for different values
of \(q\), and this range difference needs to be ignored for comparison
purposes. For \(q=1\) the index range is \([0,0.9]\). For \(q=2\) the
index range is approximately \([0, 0.5]\). We rescale the index
depending on the selected value of \(q\) against
\(I_A^{up}(q, \gamma=0.1)\), resulting in a common range \([0,1]\).

\begin{figure}

{\centering \includegraphics[width=1\linewidth]{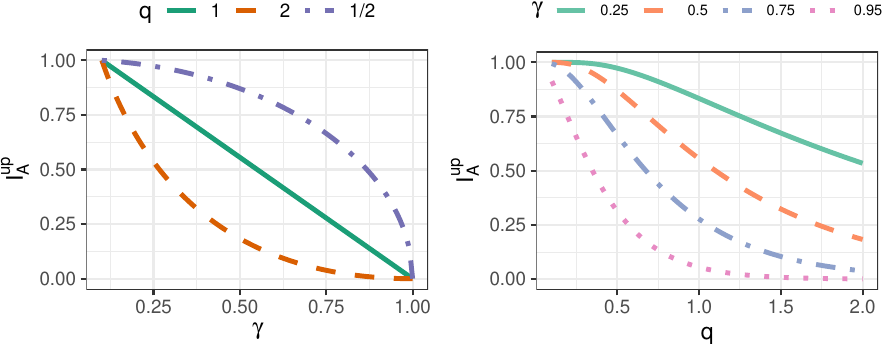} 

}

\caption{Examining the sensitivity of the index to the power parameter $q$. Rescaled values of $I_A^{up}$ as a function of $\gamma$, the fraction of non-empty bins, for fixed values of $q$ (left), and as a function of $q$ for fixed values of $\gamma$ (right). The parameter $q$ shifts emphasis between small regions and large regions of difference.}\label{fig:behavior}
\end{figure}

Fig. \ref{fig:behavior} shows the effect of \(q\) for
\(\gamma \in [0.1, 1]\), that is between 10\% and 100\% overlap in the
distributions, and the effect of \(\gamma\) for a range of
\(q \in [0.1, 1]\). When \(q=1\) the index is linearly dependent on
\(\gamma\). For \(q=2\) most of the sensitivity lies towards small
values of \(\gamma\). For \(q=1/2\) most of the sensitivity lies towards
large values of \(\gamma\). The parameter \(q\) enables the index to be
made more sensitive to small regions of difference as opposed to large
areas.

\hypertarget{visualising-the-index}{%
\subsection{\texorpdfstring{Visualising the index
\label{sec:viz}}{Visualising the index }}\label{visualising-the-index}}

A Huber plot \citep{huberpp} (available in the R package,
\texttt{PPtreeViz} \citep{pptreeviz}) is a useful illustration of a 1D
projection pursuit index. It shows the index values for a 2D
distribution across all possible 1D projections. This idea is
generalized for a 2D projection pursuit index, to illustrate the
behavior over a higher dimensional dataset, and called a \emph{topotrace
plot}. The approach is:

\begin{enumerate}
\def\labelenumi{\arabic{enumi}.}
\tightlist
\item
  Select a starting plane, one of particular interest, \(A_0\).
\item
  Randomly generate a large number (\(m\)) of directions to move away
  from the starting plane, \(A_i, i=1, ..., m\).
\item
  Generate the geodesic interpolation with a fixed length or angle
  (\(\alpha\)) in each direction,
  \(A_{ij}, i=1, ..., m; j=-\alpha, ..., 0, ..., \alpha\).
\item
  Calculate the index values along each path, \(I_{A_{ij}}\).
\item
  Plot \(I_{A_{ij}}\) against \(j\), with a separate trace for each
  \(i\).
\end{enumerate}

The purpose of making these plots is to examine the nature of the
function, in terms of local maxima and ridges, and also characteristics
such as smoothness and squint angle \citep{laa2020}. The squint angle
describes the distance from the optimal projection where a structure can
be seen -- if it is large then the function should be easier to
optimize. If the optimal projection is known, then using this as the
starting plane, provides views analogous to standing on top of the
mountain and looking down in all directions. If a random starting plane
is used, most likely this will be a low point from which to view
mountains.

\begin{figure}
\includegraphics[width=1\linewidth]{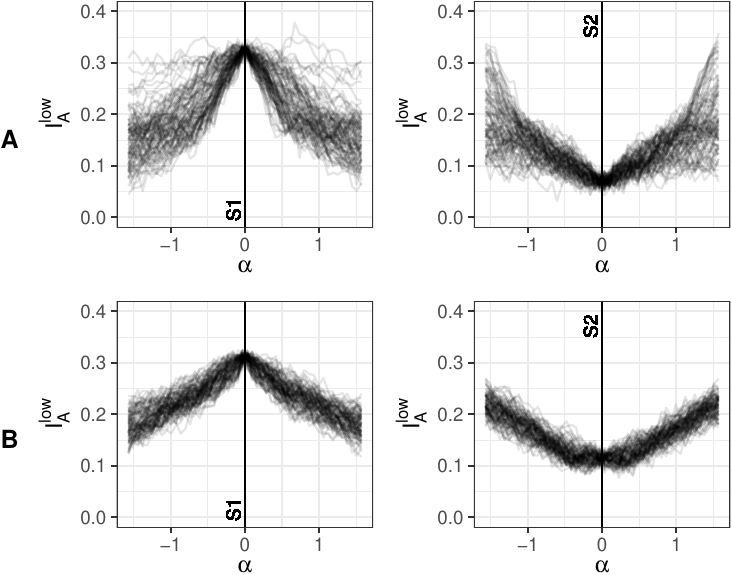} \caption{Topotrace plots showing index function characteristics for the index with $q=1$, $\alpha=\pi/2$, $m=100$, for example data sets A (top) and B (bottom) from the starting planes, the informative slice S1 (left) and the uninformative slice S2 (right). Example A indicates ridges in the function because for several traces the index remains high when varying $\alpha$. Example B has a more gradual decline from the peak. The dependence of the index on $\alpha$ is not smooth.}\label{fig:traces_q1}
\end{figure}

Fig. \ref{fig:traces_q1} shows these plots for the two example datasets
A and B, with \(\alpha = \pi/2\), \(m=100\) randomly selected
directions. The left panels show how the index value changes when moving
away from the optimal viewing slice S1, the right panel shows paths
moving away from the uninformative slice S2. We see that for the dataset
A there is a large variability between the index behavior along the
different directions. Moving away from S1 we find some flat directions
along which the index value remains large. This suggests that the
function has ridges. This is expected for this data, as a result of a
symmetry in the simulation distribution, the structure remains visible
so long as the first variable is dominant along one direction in the
plane. This also makes the structure easier to detect, and we find that
among the 100 random directions, several traces reach index values close
to that of the ideal view. On the other hand, the 100 traces shown for
set B have much less variability, and all result in an approximately
linear decay of the index value as we move away from the S1 slice.
Similarly, they all show approximately linear increase in index value
when moving away from S2. We can also see that the index is noisy,
sometimes producing jumps in index values under small rotations of the
slicing projection.

\begin{figure}
\includegraphics[width=1\linewidth]{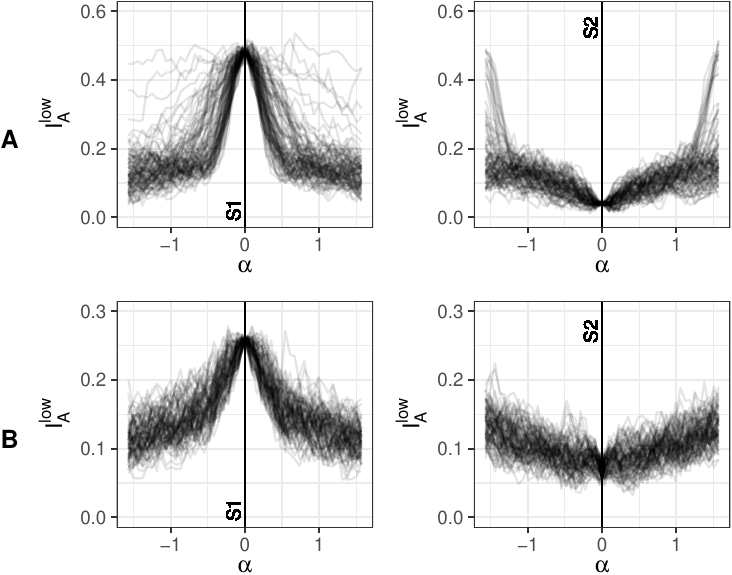} \caption{Topotrace plots for the index with $q=2$, $\alpha = \pi/2$, $m=100$ for set A (top) and set B (bottom) from the informative slice S1 (left) and the uniformative slice S2 (right). The choice of $q=2$ results in a smaller squint angle, and steeper change of the index.}\label{fig:traces_q2}
\end{figure}

We can also use this visualization to better understand the
generalizations of the index. In Fig. \ref{fig:traces_q2} we show the
topotrace plots for the same settings as Fig. \ref{fig:traces_q1}, but
with \(q=2\). We see that this results in a smaller squint angle, and
steeper change of the index near the optimal S1 slice, while the index
is flatter away from this view.

\hypertarget{index-in-practice}{%
\subsection{Index in practice}\label{index-in-practice}}

When using the index in practice we follow these steps:

\begin{enumerate}
\def\labelenumi{\arabic{enumi}.}
\tightlist
\item
  Check the underlying assumption that the data is inside a hypersphere.
  This can be ensured by first centering and scaling all variables, and
  then dropping points that have a radius above the maximum \(r_{max}\).
\item
  Choose the number of bins. Setting the bin size for polar binning uses
  the maximum radius and defines \(n_r\) equidistant radial bins and
  \(n_{\theta}\) angular bins. The preferred values will depend on the
  sample size and slice thickness.
\item
  Decide the slice radius \(h\) to use. We can think of it in terms of a
  relative resolution \(h/r_{max}\).
\item
  The binning and resolution, together with the number of variables and
  number of observations are needed to estimate the uncertainties
  following Eq.\ref{eq:eps}.
\item
  We can now define the slice index, which takes the binning and
  uncertainty estimates as input. We define the index such that for each
  slice it reweights the counts according to Eq. \ref{eq:reweight}.
\item
  For the optimization we use a modified version of the guided tour that
  passes the projected points, the distance vector and the radius \(h\)
  into the index function. This can directly be used together with the
  optimization routines from the \texttt{tourr} package.
\end{enumerate}

\hypertarget{applications}{%
\section{\texorpdfstring{Applications
\label{sec:applications}}{Applications }}\label{applications}}

\hypertarget{index-settings}{%
\subsection{Index settings}\label{index-settings}}

For the application considered we find that the following parameters
work well:

\begin{itemize}
\tightlist
\item
  We use the index as defined in Eq. \ref{eq:index}.
\item
  When binning we set \(n_r=5\) and \(n_{\theta}=8\) or \(10\). Here we
  have to consider the trade-off between resolution and required sample
  size. Note that the preferred relative resolution in radius and angle
  can be different depending on the input data and the type of
  structure. This typically requires testing different settings, similar
  to how we might look at histograms after changing the number of bins
  to understand a distribution. Ideally we select the number of bins
  that allows to resolve the features in the data without introducing
  any artifacts.
\item
  For our examples, a slice radius \(h=0.25\) worked well for scaled
  data (standard deviation = 1).
\end{itemize}

\hypertarget{classification-boundaries}{%
\subsection{Classification boundaries}\label{classification-boundaries}}

The slice display can be used to understand non-linear classification
boundaries \citep{laa2019slice}, visualized following \citet{sam.11271}
and using the \texttt{classifly} package \citep{classifly}. We use
\texttt{classifly} to sample the design space and calculate model
predictions. We can use section pursuit to identify slices that reveal
the decision boundaries in the design space, by dropping sample points
based on the assigned prediction. By selecting a class that is only
predicted in a small region we generate the hollow features to be found
by section pursuit. The resulting slice can then be viewed showing all
assigned classes to resolve the boundary. An example using the classical
olives data is discussed below, a second example studying classification
boundaries is available in the Appendix.

\hypertarget{sec:olives}{%
\subsubsection*{Olives data}\label{sec:olives}}
\addcontentsline{toc}{subsubsection}{Olives data}

We first consider the classical olives data \citep{olives}. This data
set contains measurements of 8 fatty acids for 572 Italian olive oils,
collected in 9 different areas, and is available in the
\texttt{classifly} package.

We fit a support vector machine (svm) classification model using the
implementation in the \texttt{e1071} R package \citep{e1071}, to predict
the area from the fatty acid measurements. For this example we consider
4 of the variables: palmitoleic, stearic, linoleic and arachidic, and we
use a radial kernel for the svm model. We use \texttt{classifly} for
automated sampling of the design space and evaluation of the
predictions. For the visualization we first center and scale the data to
have standard deviation one. We then select only those points that are
inside a 4D hypersphere.

\begin{figure}
\includegraphics[width=1\linewidth]{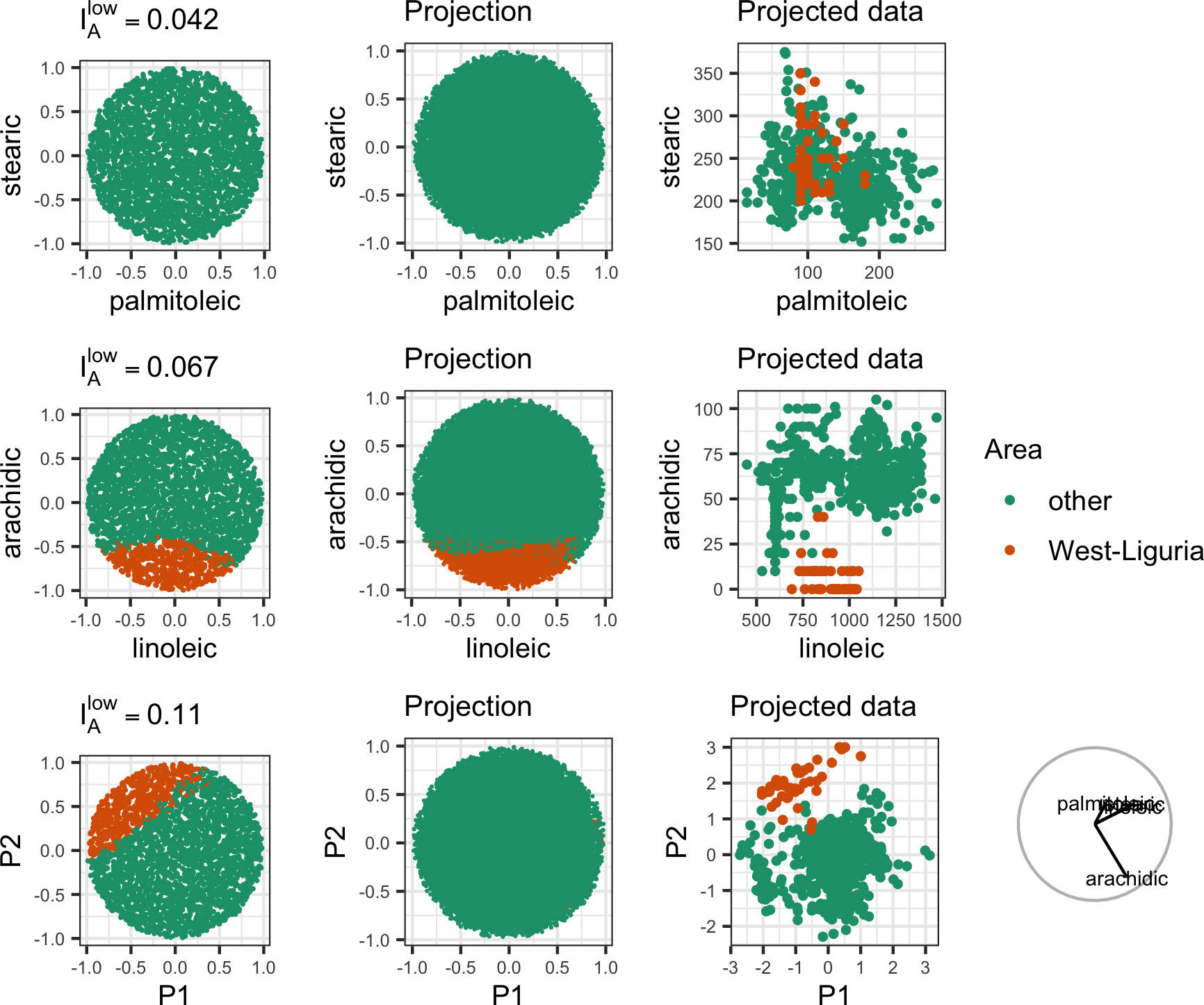} \caption{SVM classification of the West-Liguria area against all other areas in the olives data. The first and second row are projections onto pairs of variables, the last row shows the final result obtained via section pursuit. The first and second column show the svm classification in a thin slice and as a full projection, the third column shows the projected data. The last column shows guides, including the axes for the final view at the bottom.}\label{fig:olives}
\end{figure}

As an example we select as the area of interest West-Liguria and drop
all samples with this predicted class before performing section pursuit.
The model predictions are shown in slices and projections in the first
two columns of Fig. \ref{fig:olives}, the last column shows the
projected data. The views in the first row are defined by the projection
onto palmitoleic-stearic and in the second row by the projection onto
linoleic-arachidic. The color indicates if the predicted area is
West-Liguria (orange) or not (green). We see that the parameters
palmitoleic and stearic do not allow us to distinguish the West-Liguria
region in the projected data. The region where the svm model predicts
this class is hidden in the projection, and it is not predicted anywhere
inside the thin slice. On the contrary, linoleic and arachidic can
separate the West-Liguria area from the others. Looking at the model
predictions, we find that projections can partly reveal the part of the
model space resulting in this prediction, while the slice can also
resolve the non-linear decision boundary.

Next we run section pursuit on the reduced svm sample. We define the
section pursuit index with polar binning, with 5 equidistant radial bins
and 10 angular bins. The bin counts are reweighted according to Eq.
\ref{eq:reweight}. We set \(q=1\) and then optimize the index using the
simulated annealing search (\texttt{search\_better}) available in the
\texttt{tourr} package, starting form the slice defined by the
projection onto palmitoleic-stearic (see Fig. \ref{fig:olives}, first
row). The final views obtained in the optimization are shown in the
bottom row of Fig. \ref{fig:olives}.

Looking at the data projected onto linoleic and arachidic, and the
corresponding slice view of the classifier, we see a non-linear decision
boundary that allows to separate the West-Liguria area from the other
regions. This non-linearity is hidden in the projected view of the
classifier. The section pursuit has identified a slice that has a higher
index value and shows a linear decision boundary and with a larger area
of the section predicting the selected class. The projected model
predictions in this plane do not allow us to resolve this boundary.
Interestingly, to maximize the area in the sliced view of the predictor,
section pursuit has found a plane that leads to a linear decision
boundary on the projected data.

\hypertarget{inequality-condition}{%
\subsection{Inequality condition}\label{inequality-condition}}

Similar to the decision boundaries of classification models, inequality
conditions can induce non-linear boundaries in the parameter space. The
inequalities may be complicated functions of multiple parameters, and we
can use section pursuit to find visualizations that illustrate the
boundaries. These are of interest in physics models for which we often
have inequality constraints defining the allowed region of parameter
space. By understanding the shape of this region we can gain physics
insights to help guide theoretical and experimental work. For example,
the constraints might result in preferred parameter combinations hinting
at a simpler underlying model, or we can reparametrise the allowed
region to design targeted experiments to test the model further.

\hypertarget{sec:THDM}{%
\subsubsection*{Two-Higgs-Doublet Model}\label{sec:THDM}}
\addcontentsline{toc}{subsubsection}{Two-Higgs-Doublet Model}

A certain type of particle that occurs in particle physics models is the
so-called scalar (or Higgs) boson, with properties specified by a
``potential'' which depends on a set of parameters. These basic
parameters allow one to calculate physical quantities such as the masses
of the scalar particles and their interaction strengths. A common
problem that arises in this context consists of finding the parameter
space that leads to acceptable predictions satisfying known theoretical
or experimental constraints.

For example, a two-Higgs-doublet model can be written under certain
assumptions in terms of the parameter set
\(\lambda_1,\lambda_2,\lambda_3,\lambda_4,\lambda_5,\alpha,\beta\),
where the last two are angles. These angles are commonly discussed in
terms of \(\tan(\beta)\) and \(\cos(\beta-\alpha)\), as these quantities
are closely related to experimental observations. This model contains
five scalar particles (types of ``Higgs'' bosons), dubbed
\(h,H,A,H^\pm\), the first one corresponding to the famous Higgs boson
found at CERN in 2012. The squared masses of these five particles are
predicted in terms of the parameters of the model, and we use a form
from \citet{Gunion:2002zf} (given in the Appendix).

For the model to be viable it needs to satisfy a number of conditions
that restrict the parameter space that is allowed. The simplest of these
restrictions is that the masses must be real numbers: only parameters
for which all squared masses are greater than 0 are viable. We will use
this as our example, sampling
\(\lambda_1,\lambda_2,\lambda_3,\lambda_4,\lambda_5,\tan(\beta),\cos(\beta-\alpha)\)
within a 7D hypersphere. We then evaluate the predicted mass spectrum
for each sample and flag all points that result in non-physical
predictions for (some of) the masses. These points are dropped from the
dataset and we use section pursuit to find regions that lead to a
non-physical mass spectrum.

For the visualization we first standardize each parameter and then drop
all sample points for which the condition is not met. We then apply
section pursuit to the reduced sample to find sections that are
associated with real masses. We use the index with polar binning, with 5
equidistant radial bins up to the maximum radius, and 10 angular bins.
The \(\varepsilon\) cutoff is calculated according to Eq. \ref{eq:eps}
and we reweight the bins according to Eq. \ref{eq:reweight}. We set
\(q=1\) and use the \texttt{search\_better} optimization to find the
view with the maximum index value.

\begin{figure}
\includegraphics[width=1\linewidth]{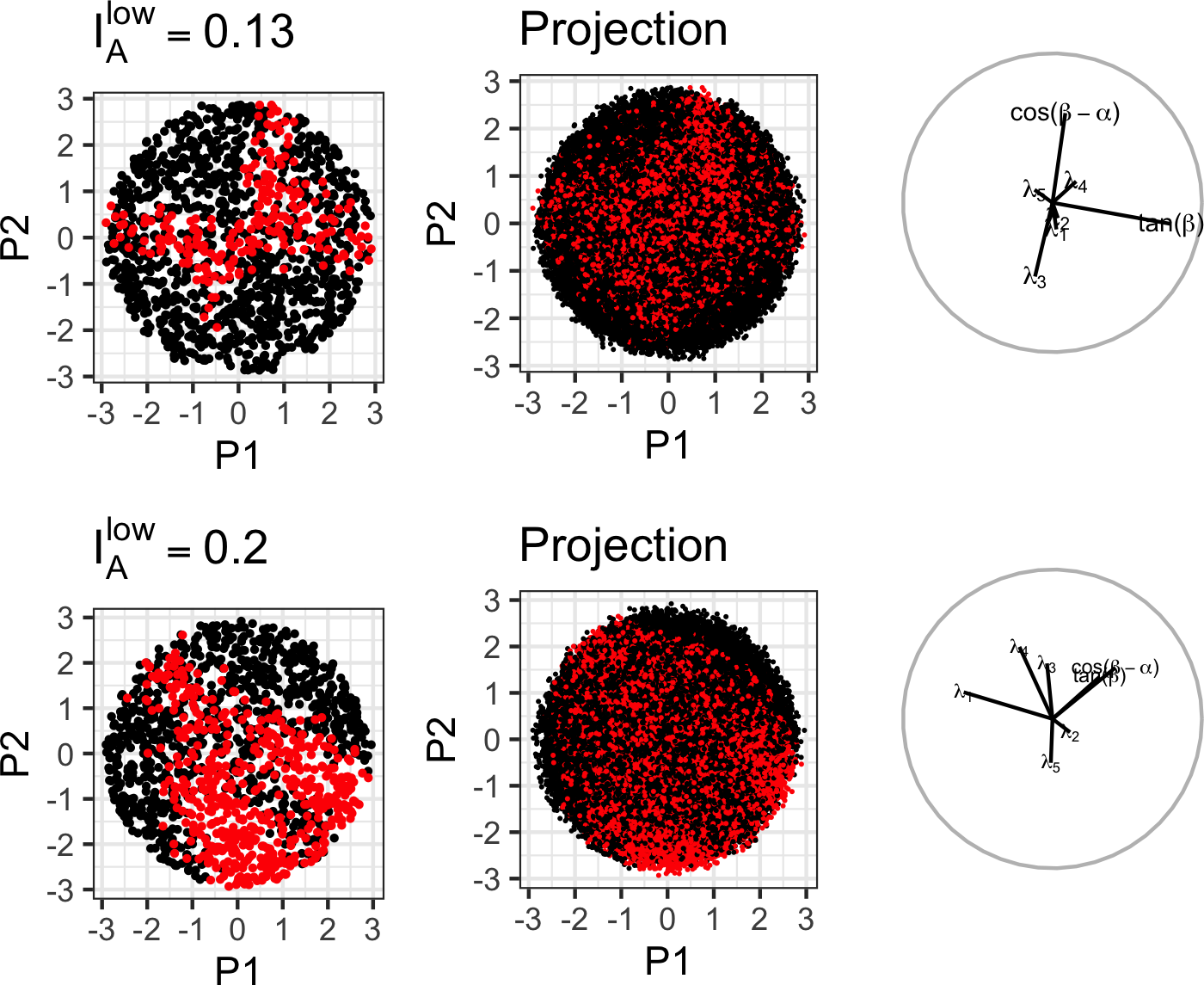} \caption{Slice (left) and projection (middle) views found when optimising the section pursuit index for the THDM sample. Black points satisfy the inequality conditions, red points do not and are dropped in the optimisation. The first row shows a view along the optimisation path, and the last row the final slice selected by section pursuit. The corresponding axes are shown on the right. The selected slices show clear separation of points not satifsying the conditions, which is hidden in the corresponding projections.}\label{fig:thdm}
\end{figure}

Two of the resulting views are shown in Fig. \ref{fig:thdm}. The samples
that violate the condition are shown in red, the remaining samples
(those used for the section pursuit) are shown in black. The top row
shows a view encountered along the optimization path, that is a section
with relatively large index value. The slice is shown on the left and
reveals the non-linear boundary defined by the conditions. The
corresponding projection is shown in the middle and cannot resolve the
feature. The bottom row shows the final view obtained via section
pursuit.

Both slices show interesting aspects of the boundary. In the first slice
the condition is only violated in a small but well-defined region of the
slice Section pursuit has identified a slice where the condition is
often violated, and the final view shows a complex non-linear boundary
between the two regions. The boundaries are hidden in the corresponding
projections. Looking at the axes representation of this projection, we
find that the optimal slice is defined by a combination of all the input
parameters, making direct interpretation challenging.

The result confirms our initial expectation that the boundaries have a
non-trivial dependence on all input parameters, and can give new
insights into its shape. To interpret it in terms of the parameters we
need to take into account both how these are contributing to the
projection, and what this means for the slice condition. Here additional
work is needed to better understand the
dependencies.\footnote{When a parameter does not contribute to the current projection, a point needs to have an observed value near the corresponding mean value to be inside the slice. The relation becomes much more complex for parameters that have a non-negligible contribution to the current projection.}

\hypertarget{conclusion-and-discussion}{%
\section{Conclusion and discussion}\label{conclusion-and-discussion}}

This paper introduces a section pursuit index that can be used to detect
hollow or dense features that are visible in slices but hidden in
projections. The index uses the distribution outside each current slice
as a reference, and computes its difference from the slice distribution.
The comparison of the two distributions is calculated as the positive
sum of differences of normalized bin counts. To avoid differences
arising from the overall shape we consider spherical multivariate
distributions. In addition, we use the expected cumulative distribution
function to reweight the inside and outside distributions separately,
such that the expected counts are uniform across all bins.

The section pursuit index can be used together with the slice display in
the \texttt{tourr} package to define a guided slice tour. This means
that we can use the available optimization routines, and we can further
look at the interpolated optimization path. In practice, to conform to
the assumption of an underlying spherical distribution, we can shave off
points outside the hypersphere. Since the focus is on detecting hidden
features in the center of the distribution, we do not expect to lose
important information doing so. In case of simulated data, augmenting
the data to be of this form is also an option.

We have shown how to use the section pursuit index to explore non-linear
decision boundaries of classification models, or similarly, to explore
complex inequality conditions that depend on multiple parameters. In all
examples, we have sampled the parameter space and evaluated the
classification or inequality conditions for all points. We then dropped
points assigned to one particular class, or in conflict with the
inequality condition to generate the hollow features in the
distribution. This allowed us to identify interesting slices with
section pursuit, and we then showed the full sample (including the
previously removed subset) with color coding in that slice to visualize
the result. For all examples considered, the optimization resulted in
slice views that illustrate interesting aspects of the boundaries.
Additional potential applications for section pursuit include the
exploration of non-standard multivariate confidence regions or Bayesian
credible regions.

Note that large samples are required to obtain useful insights using the
section pursuit method. As discussed in the context of sufficient sample
size, the number of data points quickly grows with the number of
dimensions. The applications we have studied here are considering
simulated data in up to seven dimensions. In the case of larger
parameter space we recommend to first reduce dimensionality: in the
olives data example we have selected variables based on prior knowledge,
in the PDFSense example in the Appendix we have used principal component
analysis prior to the application of section pursuit. The method is not
appropriate for small sample size data, in those cases interactive
methods such as linked brushing will be preferred.

Depending on the application, different index definitions would be
recommended. Several variations on the index would be produced through
different choices of weights, and parameters in the generalized index
definition. This paper hasn't fully explored the impact of all the
choices, but the same type of diagnostics, in particular, the index
visualization presented in Section \ref{sec:viz}, could be used decide.

Given the compositional nature of the normalized bin-count data,
considering e.g.~log-ratios in the definition, could also define an
interesting index. It may also be reasonable to consider using a kernel,
instead of discrete bins, to produce some spatial smoothing. The effect
would be to focus attention on large differences in specific regions
rather than small differences anywhere. A completely different approach
to producing a section index, would be to apply any existing projection
pursuit index on a sliced projection instead of the full projected data.
Only observations inside the slice are then used for the index
calculation.

\hypertarget{acknowledgements}{%
\section*{Acknowledgements}\label{acknowledgements}}
\addcontentsline{toc}{section}{Acknowledgements}

The authors gratefully acknowledge the support of the Australian
Research Council. This article was created with \texttt{knitr}
\citep{knitr} and R Markdown \citep{rmarkdown} with embedded code, using
the \texttt{tidyverse} \citep{tidyverse} packages. We thank the Wharton
Statistics Department at the University of Pennsylvania for their
hospitality while part of this work was conducted and Buja was on their
faculty. Open access funding provided by University of Natural Resources
and Life Sciences Vienna (BOKU).

\hypertarget{supplementary-material}{%
\section*{Supplementary material}\label{supplementary-material}}
\addcontentsline{toc}{section}{Supplementary material}

\begin{itemize}
\tightlist
\item
  Code and data is available at
  \url{https://github.com/uschiLaa/paper-section-pursuit}.
\item
  The Appendix contains the derivation of the radial CDF of a
  hypersphere projected onto a 2D plane and the equations used to
  calculate the masses in the two-Higgs-doublet model, and an additional
  example from physics.
\end{itemize}

\bibliographystyle{tfcad}
\bibliography{interactcadsample.bib}



\section*{Supplementary material}

# Appendix: Hole or grain? A Section Pursuit Index for Finding Hidden Structure in Multiple Dimensions

## Slice volume derivation

The relative volume of a slice through a $p$ dimensional hypersphere of radius $R$ was previously derived in Laa, Cook, and Valencia (2019). Below we reproduce the calculation for the convenience of the reader. We start with the volume of the hypersphere with radius $r$ in $q$ dimensions:

$$V(r, q) = \frac{\pi^{q/2} r^q}{\frac{q}{2} \Gamma(q/2)}. \tag{1}$$

The variation of the volume with the radius is given as

$$\frac{dV(r, q)}{dr} = 2 \frac{\pi^{q/2} r^{q-1}}{\Gamma(q/2)}. \tag{2}$$

The volume in a slice is spherical in the orthogonal space ($p-2$ dimensions) and capturing the full area within the plane. We thus compute it by integrating the product of $\frac{dV(r,p-2)}{dr}$ and the area in the plane parametrised by $r$. The area in the plane is a circle with radius $\sqrt{R^2 - r^2}$. The slice volume is thus calculated as

$$V_{slice}(R, h, p) = \int_0^h \frac{dV(r, p-2)}{dr} V(\sqrt{R^2 - r^2}, 2) dr = \frac{\pi^{p/2}}{\Gamma(p/2)} \frac{h^{p-2}}{p} [pR^2 - (p-2)h^2]. \tag{3}$$

The relative volume of the slice given by the fraction

$$V_{rel}(R, h, p) = \frac{V_{slice}(R, h, p)}{V(R, p)} = \frac{h^{p-2}}{2R^p} [pR^2 - (p-2)h^2]. \tag{4}$$

## Simulated data

The four-dimensional hyperspherical harmonics (Domokos 1967) can be written as

$$Z_{n\ell}^m(\beta, \theta, \phi) = 2^{\ell+1/2} \sqrt{\frac{(n+1)\Gamma(n-\ell+1)}{\pi \Gamma(n+\ell+2)}} \Gamma(\ell+1) \sin^\ell \beta C_{n-\ell}^{(\ell+1)}(\cos \beta) Y_\ell^m(\theta, \phi), \tag{5}$$

where $C_{n-\ell}^{(\ell+1)}$ are the Gegenbauer polynomials and $Y_\ell^m(\theta, \phi)$ are the 3D spherical harmonics. The polynomials are defined over the 4D unit hypersphere $S^3$ in $\mathbb{R}^4$ and parameterized by three angles: the azimuthal angles $\phi$, the 3D zenith (or polar) angle $\theta$ and the 4D zenith angle $\beta$.

The data sets (A, B) are generated from two of these polynomials. Explicitly they are obtained from,

$$
\begin{aligned}
\text{Set A}: Z_{20}^0 &= \frac{3 - 4\sin^2 \beta}{\sqrt{2}\pi} \\
\text{Set B}: \sqrt{\frac{3}{2}} Z_{53}^1 &= \sqrt{\frac{3}{10}} \frac{1}{2\pi} e^{i\phi} \sin^3 \beta \left(40 \cos^2 \beta - 4\right) \sin \theta \left(1 - 5 \cos^2 \theta\right)
\end{aligned} \tag{6}
$$

To generate the data sets we start from a uniform sample of points inside a 4-ball of radius one (the inside of the hypersphere $S^3$) which we then classify into inside or outside points depending on whether $r$ is

1

less than or greater than $|Z_{n\ell}^m(\beta, \theta, \phi)|$. This procedure generates a much larger number of outside than inside points so we collect them separately to obtain reasonably sized samples. The uniform sampling is done in Cartesian coordinates $(x, y, z, t)$ which are then transformed to hyper-spherical coordinates $(r \cos \beta, r \sin \beta \cos \theta, r \sin \beta \sin \theta \cos \phi, r \sin \beta \sin \theta \sin \phi)$.

## Radial CDF of projected hyperspheres

The radial CDF used throughout this work can be derived by calculating the fraction of the projected volume within a circle of radius $r$. The volume of a $p$ dimensional hypersphere with radius $R$ is

$$V(p, R) = \frac{2\pi^{p/2} R^p}{p \Gamma(p/2)} \tag{7}$$

and the projected volume *outside* a circle of radius $r$ is

$$V_{outside}(r, p, R) = \int_r^R V(p - 2, \sqrt{R^2 - \rho^2}) 2\pi \rho d\rho = \frac{\pi^{p/2}(R^4 - r^2 R^2)^{p/2}}{R^p \Gamma(p/2 + 1)}. \tag{8}$$

The projected volume *inside* the circle is given by

$$V_{inside}(r, p, R) = V(p, R) - V_{outside}(r, p, R), \tag{9}$$

and we therefore get the relative volume within the circle (and thus the radial CDF) as

$$F(r, p, R) = \frac{V_{inside}(r, p, R)}{V(p, R)} = 1 - \left(1 - \left(\frac{r}{R}\right)^2\right)^{p/2}. \tag{10}$$

## Squared masses in the two-Higgs-doublet model

Following Gunion and Haber (2003) we compute the squared masses as

$$
\begin{aligned}
m_h^2 &= \frac{v^2}{\sin(\beta - \alpha)} \left(-\lambda_1 \cos^3 \beta \, \sin \alpha + \lambda_2 \sin^3 \beta \cos \alpha + \frac{\lambda_{345}}{2} \, \cos(\beta + \alpha) \sin(2\beta)\right) \\
m_H^2 &= \frac{v^2}{\cos(\beta - \alpha)} \left(\lambda_1 \cos^3 \beta \, \cos \alpha + \lambda_2 \sin^3 \beta \sin \alpha + \frac{\lambda_{345}}{2} \, \sin(\beta + \alpha) \sin(2\beta)\right) \\
m_{H^\pm}^2 &= \frac{v^2}{\sin(2(\beta - \alpha))} \left(-\sin(2\alpha) \, (\lambda_1 \cos^2 \beta - \lambda_2 \, \sin^2 \beta) + \lambda_{345} \, \sin(2\beta) \cos(2\alpha) - \frac{\lambda_{45}}{2} \, \sin(2(\beta - \alpha))\right) \\
m_A^2 &= \frac{v^2}{\sin(2(\beta - \alpha))} \left(\sin(2\alpha) \, (-\lambda_1 \cos^2 \beta + \lambda_2 \, \sin^2 \beta) + \lambda_{345} \, \sin(2\beta) \cos(2\alpha) - \lambda_5 \sin(2(\beta - \alpha))\right)
\end{aligned}
\tag{11}
$$

where we have used the shorthand notation $\lambda_{ij\ldots} = \lambda_i + \lambda_j + \cdots$ and the constant $v \approx 246$~GeV sets the scale for the masses.

## Additional application: PDFSense data

As similar example, but with different behavior, is given by the decision boundaries of a classification model for the PDFSense dataset. This data has 4021 observations in a 56 dimensional parameter space, that are grouped into 3 classes (Wang et al. 2018). Following the analysis in Cook, Laa, and Valencia (2018) we only consider the first six principal components, and train an svm classification model with radial kernel. We again use `classifly` and select the resulting sample points inside a 6D hypersphere after individual centering and scaling each of the variables.

For this example, the classifier wraps tightly around two of the groups (DIS and jets), and most of the space is filled by the third group (VBP). We select only the samples predicting the VBP class, generating a sample

with a small hollow region to be found by section pursuit. We again use the index with polar binning, with 5 equidistant radial bins and 8 angular bins. We set $q = 1$ and use the `tourr` function `search_better` to find the view with the maximum index value.

The result is shown in Fig. 1, comparing the starting projection (top row) and the final view (bottom row). In the final slice view of the classifier predictions we see the decision boundaries between the three classes. The different regions are hidden in the projection of the classifier and of the data. In this case the svm model uses additional information from the orthogonal directions to separate the three classes. By looking at the final slice from section pursuit we have obtained a conditional view that allows us to resolve the resulting boundary.

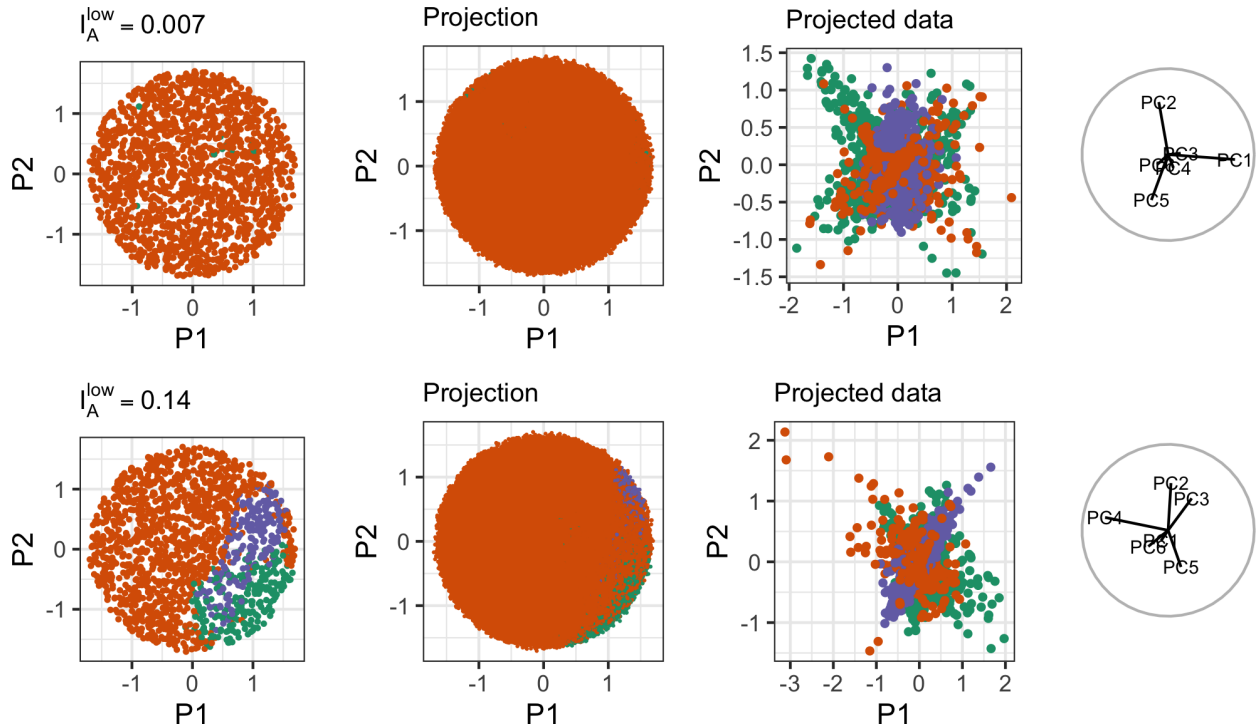

Figure 1: SVM classification of the PDFSense data, with predicted classes mapped to color. The first row shows a random starting projection, the second row is the final projection obtained via section pursuit on the second class shown in orange. The first and second column show the predicted class label from the svm in a thin slice and a projection. The third column shows the projected data in the same plane, and the last column shows the axes of the corresponding projection.



\end{document}